\documentclass[twocolumn,aps,prc,superscriptaddress,nofootinbib,floatfix,longbibliography]{revtex4}
\usepackage{url}
\usepackage{cancel}
\usepackage[colorlinks,linkcolor=blue,citecolor=blue,filecolor=black,urlcolor=blue]{hyperref}
\usepackage{epsfig,graphics}
\usepackage{graphicx}
\usepackage{dcolumn}
\usepackage{bm}
\usepackage{amssymb}
\usepackage{amsmath}
\usepackage{multirow}
\usepackage{float}
\usepackage{harpoon}
\usepackage{MnSymbol}
\usepackage{appendix}
\usepackage{color}
\usepackage{hyperref}
\usepackage{cleveref}
\usepackage{CJKutf8}

\newcommand{\snn}{\mbox{$\sqrt{s_{\mathrm{NN}}}$}}
\newcommand{\pT} {p_{\mathrm{T}}}

\newcommand{\lr}[1]{\left\langle #1\right\rangle}

\newcommand{\pa} {p_{\mathrm{T1}}}
\newcommand{\pb} {p_{\mathrm{T2}}}

\newcommand{\pTc} {p_{\mathrm{T}}^{0}}
\newcommand{\pTa} {p_{\mathrm{T}}^{\mathrm{A}}}

\graphicspath{{./}{../}}

\begin{document}
\title{Sources of Radial Flow Fluctuations in the Quark-Gluon Plasma}
\newcommand{\bnl}{Physics Department, Brookhaven National Laboratory, Upton, NY 11976, USA}
\newcommand{\sbu}{Department of Chemistry, Stony Brook University, Stony Brook, NY 11794, USA}
\author{Jiangyong Jia (\begin{CJK*}{UTF8}{gbsn}贾江涌\end{CJK*})}\affiliation{\sbu}\affiliation{\bnl}
\date{\today}
\begin{abstract}
The differential radial flow fluctuation $v_0(p_{\mathrm{T}})$ has emerged as a new probe of the quark-gluon plasma. However, its characteristic rise-and-fall pattern with $\pT$, resembling anisotropic flow, remains unexplained. I introduce a momentum rescaling framework that factorizes $v_0(p_{\mathrm{T}})$ into kinematic and dynamical components: $v_0(p_{\mathrm{T}})/v_0 = -[d\ln\langle n(p_{\mathrm{T}})\rangle/d\ln p_{\mathrm{T}} + 1] \times g(p_{\mathrm{T}})$. The first factor, determined by spectral shape, generates the rise-and-fall pattern as the spectra transition from exponential to power-law behavior. The dynamical component $g(p_{\mathrm{T}})$ isolates $p_{\mathrm{T}}$-dependent dynamics: $<1$ signals suppressed fluctuations, $>1$ indicates enhancement. Analysis of LHC data reveals $g(p_{\mathrm{T}})$ deviates from unity by 20--40\% in central collisions. Predictions for RHIC show that spectral shape alone generates the rise-and-fall baseline pattern with substantial energy dependence. This framework enables tighter medium property constraints by separating kinematic from dynamical effects, with broad applications to anisotropic flow and higher-order radial flow fluctuations.
\end{abstract}
\maketitle

\begin{figure}[htbp!]
\centering
 \includegraphics[width=1\linewidth]{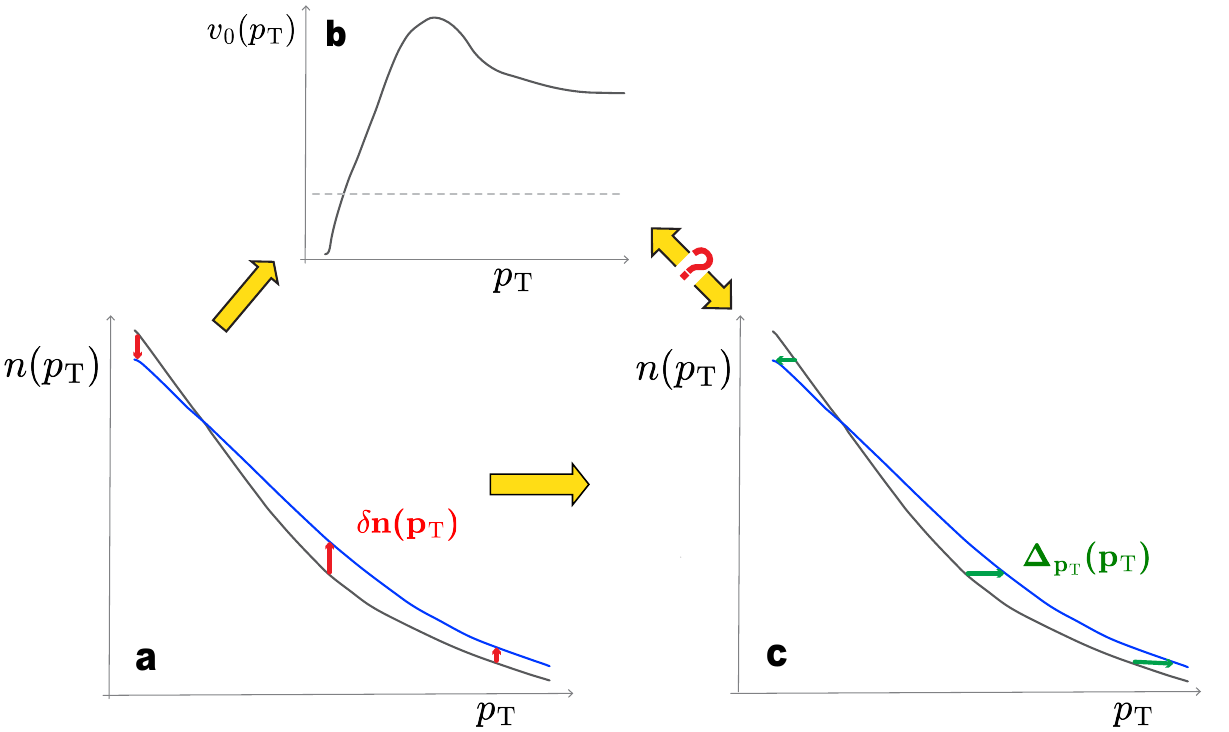}
\vspace*{-0.5cm}\caption{\textbf{Cartoon illustration of the momentum rescaling model and its connection to $v_0(\pT)$.} (\textbf{a}) Shows the global event-by-event fluctuation of fractional spectra yield, $\delta n(\pT)/\lr{n(\pT)}$ with $\int n(\pT)d\pT=1$. The blue curve indicates the change of $n(\pT)$ in an event with large $[\pT]$ from the average $\lr{n(\pT)}$ represented by the black curve. (\textbf{b}) Shows how these fluctuations can be mapped to the characteristic shape of the $v_0(\pT)$~\cite{ATLAS:2025ztg}, which typically rises from negative values, crosses zero, peaks, and then falls. (\textbf{c}) Depicts the core concept of our model: these fluctuations are interpreted as a rescaling of the $\pT$-axis, $\Delta_{\pT} (\pT)/\pT = g(\pT) \delta[\pT]/\lr{[\pT]}$. Our primary goal is to extract this effective $\pT$-dependent momentum scaling factor, $g(\pT)$, from the experimentally measured $v_0(\pT)$ and average spectrum $\lr{n(\pT)}$.}
\label{fig:1}
\vspace*{-0.2cm}
\end{figure}

{\bf Introduction.} Heavy ion collisions at RHIC and the LHC create the quark-gluon plasma (QGP), a state of matter composed of deconfined quarks and gluons. A defining feature of QGP dynamics is radial flow, the isotropic collective expansion driven by pressure gradients that develop in the collision's earliest moments~\cite{Heinz:2013th}. This radial flow manifests directly in the transverse momentum ($\pT$) spectrum of produced particles, $n(\pT)$: stronger flow creates flatter spectra and higher average momentum $[\pT]$, while weaker flow yields steeper spectra (Fig.~\ref{fig:1}a). 

The strength of radial flow correlates with the initial collision geometry. Even at a given centrality, events exhibit significant fluctuations in their overall sizes. These size variations translate into flow variations: smaller, denser initial configurations generate stronger collective expansion than larger, more dilute ones~\cite{Bozek:2012fw,Giacalone:2020dln,ATLAS:2024jvf}.

Event-by-event fluctuations in radial flow are characterized by two complementary observables. The $\pT$-integrated measure, $v_0 \equiv \sqrt{\langle(\delta [\pT])^2\rangle}/\langle[\pT]\rangle$, where $\delta[\pT] = [\pT] - \langle [\pT] \rangle$~\footnote{We use ``[ ]'' to denote average of a quantity in a given event, and $\lr{}$ to denote the average over an event ensemble.}, quantifies the overall magnitude of momentum fluctuations. The $\pT$-differential measure quantifies how spectral fluctuations $\delta n(\pT) = n(\pT) - \langle n(\pT) \rangle$ correlate with momentum fluctuations~\cite{Mazeliauskas:2015efa,Schenke:2020uqq}:
\begin{align}\label{eq:1}
v_0(\pT) = \rho(n(\pT),[\pT]) \frac{\sqrt{\lr{(\delta n(\pT))^2}}}{\lr{n(\pT)}}\;,
\end{align}
where $\rho(x,y) = \lr{\delta x \delta y}/\sqrt{\lr{(\delta x)^2}\lr{(\delta y)^2}}\in [-1,1]$ is the Pearson correlation coefficient. Since $\rho$ can be negative, $v_0(\pT)$ takes negative values at low $\pT$ where spectra steepen with increasing flow. The power of $v_0(\pT)$ lies in its ability to distinguish between different radial flow modes---subleading density profiles that produce identical average momenta but generate distinct spectral fluctuation patterns~\cite{Mazeliauskas:2015efa}.

While both observables are sensitive to initial conditions and medium properties such as the equation of state and viscosities, their ratio $v_0(\pT)/v_0$ removes overall magnitude variations, isolating the $\pT$-dependent physics. This normalized observable has emerged as a particularly promising probe of bulk viscosity, being largely insensitive to shear viscosity and the equation of state~\cite{Samanta:2023amp,Du:2025dpu}.

Recent measurements by ATLAS~\cite{ATLAS:2025ztg} and ALICE~\cite{ALICE:2025iud} have revealed a universal pattern in $v_0(\pT)$: it rises from negative values at low $\pT$, crosses zero near the average momentum, peaks around 3--4 GeV, and then decreases at higher momenta (Fig.~\ref{fig:1}b). This characteristic rise-and-fall shape bears superficial resemblance to anisotropic flow coefficients $v_n(\pT)$~\cite{ATLAS:2012at}, where the rising portion reflects collective expansion and the falling portion signals jet quenching. However, this analogy may be problematic for $v_0(\pT)$ since this pattern persists even in peripheral collisions where jet quenching should be negligible. What, then, is the physical origin of this universal shape?

The data also present a puzzling dichotomy. Normalizing $v_0(\pT)$ by the integral $v_0$ dramatically reduces the centrality dependence at low $\pT$ ($\pT < 2$ GeV), suggesting a universal underlying mechanism. Yet significant variations persist at higher momenta ($2 < \pT < 4$ GeV)~\cite{ATLAS:2025ztg}. These high-$\pT$ variations have been tentatively attributed to bulk viscosity effects or subleading radial flow modes~\cite{Parida:2024ckk}. A key question is: what would $v_0(\pT)/v_0$ look like in the absence of these effects? Without establishing such a baseline, interpreting centrality-dependent deviations remains ambiguous. 

This Letter addresses these challenges through an analytical framework that connects $v_0(\pT)$ to its underlying physical origin. Since radial flow boosts particles toward higher momentum, event-by-event spectral fluctuations can be recast as a $p_{\mathrm{T}}$-dependent momentum rescaling transformation: $\Delta_{\pT}(\pT)/\pT = g(\pT) \delta[\pT]/\lr{[\pT]}$, where $g(\pT)$ is a $\pT$-dependent scaling function (Fig.~\ref{fig:1}c).

This approach factorizes $v_0(\pT)$ into two distinct components: a kinematic component determined purely by the spectral shape, and a dynamical component $g(\pT)$ that encodes deviations from global momentum rescaling. Comparison of our predictions with ATLAS data shows that the kinematic component captures much of the observed features of $v_0(\pT)/v_0$. However, significant dynamical contributions with clear centrality dependence remain. By establishing the spectral-shape baseline, we can cleanly extract the physically meaningful $g(\pT)$ that reveals underlying dynamics such as bulk viscosity, jet quenching, or subleading flow modes.

{\bf Momentum rescaling model.} 
Eq.~\eqref{eq:1} defines $v_0(\pT)$ operationally. Its physical content becomes transparent when spectral fluctuations are assumed to be driven by a single collective variable, $[\pT]$~\cite{Parida:2024ckk} (Fig.~\ref{fig:1}a):
\begin{align}\label{eq:2}
\frac{\delta n(\pT)}{\lr{n(\pT)}} =\frac{v_0(\pT)}{v_0}\frac{\delta [\pT]}{\lr{[\pT]}}\;,
\end{align}
This single-mode ansatz is consistent with Eq.~\eqref{eq:1} and is supported by the factorization $v_0(\pa)v_0(\pb)$ observed experimentally~\cite{ATLAS:2025ztg}.

We model the normalized single-event spectrum, $n(\pT)$ with $\int n(\pT) d\pT=1$. Assuming that fluctuations around the event-averaged spectrum $f(\pT) \equiv \lr{n(\pT)}$ arise from variations in $[\pT]$, we can produce event-by-event spectra by rescaling the momentum axis:
\begin{align}\label{eq:3}
n(\pT) = \frac{\lr{[\pT]}}{[\pT]} f\left(\pT\frac{\lr{[\pT]}}{[\pT]}\right)\;,
\end{align}
where the prefactor ensures particle number conservation. For small fluctuations, Taylor expansion yields:
\begin{align}\label{eq:4}
\frac{\delta n(\pT)}{\lr{n(\pT)}} \approx \left.\frac{\partial\ln \lr{n(\pT)}}{\partial\ln [\pT]}\right|_{[\pT]=\lr{[\pT]}} \frac{\delta [\pT]}{\lr{[\pT]}}\;. 
\end{align}
Comparing with Eq.~\eqref{eq:2} gives our central relation:
\begin{align}\label{eq:5}
\frac{v_0(\pT)}{v_0} &= \left.\frac{\partial\ln \lr{n}}{\partial\ln [\pT]}\right|_{[\pT]=\lr{[\pT]}}= -\frac{d\ln \lr{n(\pT)}}{d\ln\pT}-1\;,
\end{align}
where the $-1$ term arises from the particle number conservation constraint $\int n(\pT) d\pT = 1$, which imposes the sum rule $\int v_0(\pT) \lr{n(\pT)} d\pT = 0$~\cite{Parida:2024ckk}. Without it, the spectral index $-d\ln\lr{n}/d\ln\pT$ is strictly positive for any falling spectrum and $v_0(\pT)$ would never cross zero. The existence and range-dependence of the zero-crossing point is thus imposed by the normalization convention rather than by the underlying dynamics~\cite{Bhatta:2025oyp}.

Under constant fractional momentum rescaling, $v_0(\pT)$ is determined entirely by spectral shape. For an exponential spectrum characteristic of thermal emission, $\lr{n(\pT)} \propto \pT \exp(-2\pT/[\pT])$, we recover~\cite{Schenke:2020uqq}
\begin{align}\label{eq:5a}
v_0(\pT)/v_0 = 2(\pT/\lr{[\pT]}-1)\;.
\end{align}
For a power-law spectrum at high $\pT$, $\lr{n(\pT)}\propto (\pT+p_0)^{-m}$, the result becomes
\begin{align}\label{eq:5b}
v_0(\pT)/v_0 = \frac{m\pT}{(\pT+p_0)}-1 \underset{\pT \to \infty}{=\!=}  m-1\;,
\end{align}
where $m$ is the spectral power index. The transition of spectral shape between these regimes at $\pT\approx3$--4 GeV naturally produces the observed rise-and-fall pattern.

To capture deviations from global rescaling, a $\pT$-dependent factor is introduced:
\begin{align}\label{eq:6}
\frac{v_0(\pT)}{v_0}= -g(\pT)\left(\frac{d\ln\lr{n(\pT)}}{d\ln\pT}+1\right)\;.
\end{align}
Values $g(\pT)\neq1$ signal additional $\pT$-dependent dynamics beyond pure momentum rescaling.

{\bf Implementation.} 
Extracting $g(\pT)$ from experimental data follows these steps:
\begin{enumerate}
\item {\it Baseline calculation:} Compute the expected $v_0(\pT)$ assuming pure global rescaling ($g(\pT)=1$) using published spectra via Eq.~\eqref{eq:5}.

\item {\it Acceptance correction:} The $v_0$ measured within a limited range $\pT\in A$ relates to the full-acceptance value via~\cite{Schenke:2020uqq,Parida:2024ckk}:
\small{\begin{align}\label{eq:8}
\hspace*{0.4cm}C_A = \frac{v_0^A}{v_0} = \frac{\int_A (\pT - \lr{[\pT]}_A) \lr{n(\pT)} v_0(\pT)/v_0 d\pT} {\int_A \pT\lr{n(\pT)} d\pT}\;,
\end{align}}\normalsize
yielding the model prediction: 
\begin{align}\label{eq:9}
\left( \frac{v_0(\pT)}{v_0^A} \right)_{\text{model}} = - \left(\frac{d \ln \lr{n(\pT)}}{d \ln \pT}+1\right )/C_A\;. 
\end{align}

\item  {\it Offset correction:} The model uses the full-range spectrum while experimental $v_0(\pT)$ is normalized in a restricted range, introducing a constant vertical offset that carries no dynamical information~\cite{Bhatta:2025oyp}. We align the zero-crossing points:
\begin{align}\label{eq:10}
\hspace*{0.4cm} s = \left( \frac{v_0(\pTc)}{v_0^A} \right)_{\text{model}}- \left( \frac{v_0(\pTc)}{v_0^A} \right)_{\text{data}}\!,
\end{align}
where $\pTc$ is the $\pT$ at which $v_0(\pTc)= 0$ in the data. This offset vanishes when identical normalization ranges are used.

\item {\it Extract $g(\pT)$:} A ratio between measured~\cite{ATLAS:2025ztg} and offset-corrected model values yields:
\begin{align}\label{eq:11}
g(\pT) = \frac{ (v_0(\pT)/v_0^A)_{\text{data}}}{ -(d\ln \lr{n(\pT)}/d \ln \pT + 1)/C_A -s}\;.
\end{align}
The differential momentum fluctuation shown in Fig.~\ref{fig:1}c follows as:
\begin{align}\label{eq:12}
\frac{\Delta_{\pT}(\pT)}{\pT} \equiv g(\pT) v_0 =  g(\pT) \frac{v_{0,\textrm {data}}^A}{C_A}\;.
\end{align}
\end{enumerate}

{\bf Results.} We use charged particle spectra from ALICE~\cite{ALICE:2018vuu} and ATLAS~\cite{ATLAS:2022kqu} for $pp$ and Pb+Pb collisions at $\snn=5.02$ TeV, spanning 0.15 to $>$100 GeV. The corresponding $v_0$ and $v_0(\pT)$ measurements are from ATLAS~\cite{ATLAS:2025ztg}.

\begin{figure}[h!]
    \centering
 \includegraphics[width=1\linewidth]{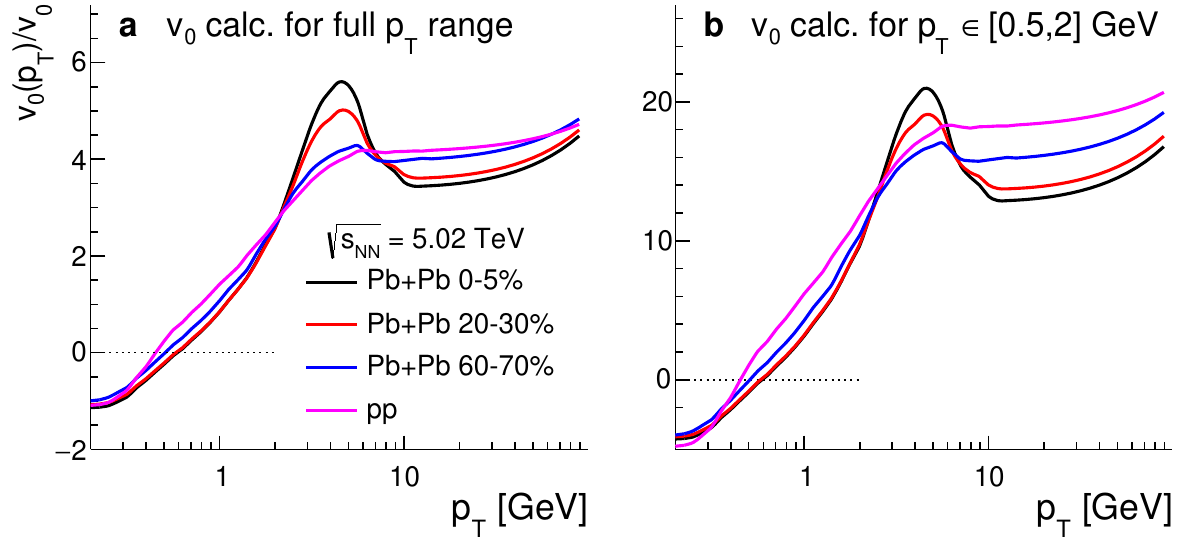}
    \caption{The $v_0(\pT)/v_0$ (\textbf{a}) and $v_0(\pT)/v_0^A$ (\textbf{b}) calculated from charged hadron spectra via Eq.~\eqref{eq:5} and Eq.~\eqref{eq:9}, respectively, assuming constant fractional momentum rescaling ($g(\pT)=1$). Results are shown for $pp$ and several centrality classes in Pb+Pb collisions at $\snn = 5.02$ TeV. The spectral uncertainties influencing $v_0(\pT)/v_0$ are very small and hence not included.}
    \label{fig:2}
\end{figure}

Figure~\ref{fig:2}a demonstrates that global rescaling ($g(\pT)=1$) reproduces the characteristic rise-and-fall pattern in central collisions. This signature weakens progressively toward peripheral and $pp$ collisions, consistent with expectations.
\begin{figure*}[htbp]
    \centering
 \includegraphics[width=0.8\linewidth]{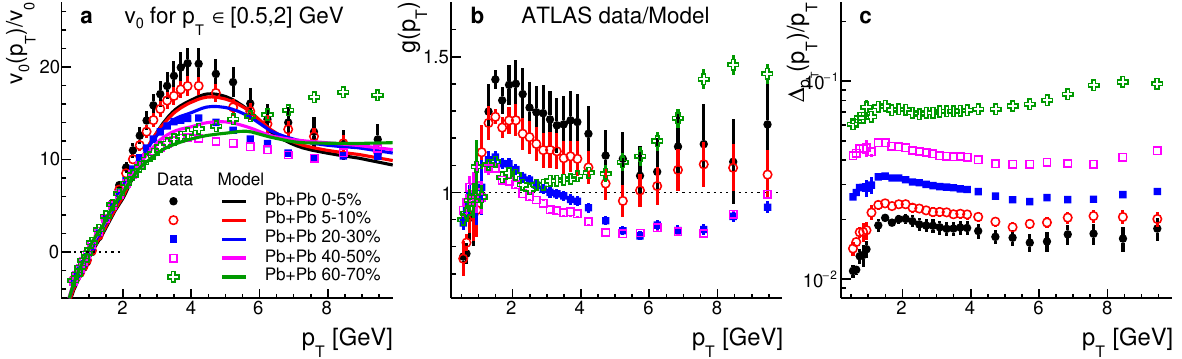}
\vspace*{-0.3cm}
    \caption{\textbf{Extracting physics from ATLAS $v_0(\pT)$ data.} (\textbf{a}) Calculated $v_0(\pT)/v_0^A$ (lines) after zero-crossing alignment (Eq.~\eqref{eq:10}), compared with ATLAS data (markers) for Pb+Pb centrality classes at $\snn=5.02$ TeV. (\textbf{b}) Extracted $g(\pT)$ (markers) via Eq.~\eqref{eq:11}. The dashed line at $g(\pT)=1$ indicates global momentum rescaling. (\textbf{c}) Differential momentum rescaling factor $\Delta_{\pT}(\pT)/\pT$ from Eq.~\eqref{eq:12}.}
\label{fig:3}
\vspace*{-0.3cm}
\end{figure*}

The acceptance factor $C_A$ for ATLAS's reference range $A=[0.5,2]$ GeV varies from 0.228 (peripheral) to 0.269 (central), as detailed in Appendix~\ref{app:1}. Consequently, this correction increases $v_0(\pT)/v_0^A$ relative to $v_0(\pT)/v_0$ (Fig.~\ref{fig:2}b).

Figure~\ref{fig:3}a compares model predictions with ATLAS data, revealing significant deviations at intermediate $\pT$. The extracted $g(\pT)$ (Fig.~\ref{fig:3}b) shows systematic departures from unity, most pronounced in central collisions: $g(\pT)\sim0.8$ at low $\pT$, peaking near 1.4 around 2 GeV, then decreasing to $\sim$1.2 at high $\pT$. Peripheral collisions show substantially weaker deviations.

These departures from $g(\pT)=1$ may arise from: (1) $\pT$-dependent viscous corrections modifying flow profiles, (2) resonance decays and hadronic rescattering altering momentum distributions, or (3) jet production and quenching creating fluctuations uncorrelated with low-$\pT$ radial flow. The strong centrality dependence indicates these effects vary with system size and require further theoretical investigation.

Figure~\ref{fig:3}c presents $\Delta_{\pT}(\pT)/\pT$, showing mild $\pT$ dependence but strong enhancement toward peripheral collisions, primarily reflecting the centrality-dependent integral $v_0$.

\begin{figure}[h!]
 \centering
 \includegraphics[width=0.9\linewidth]{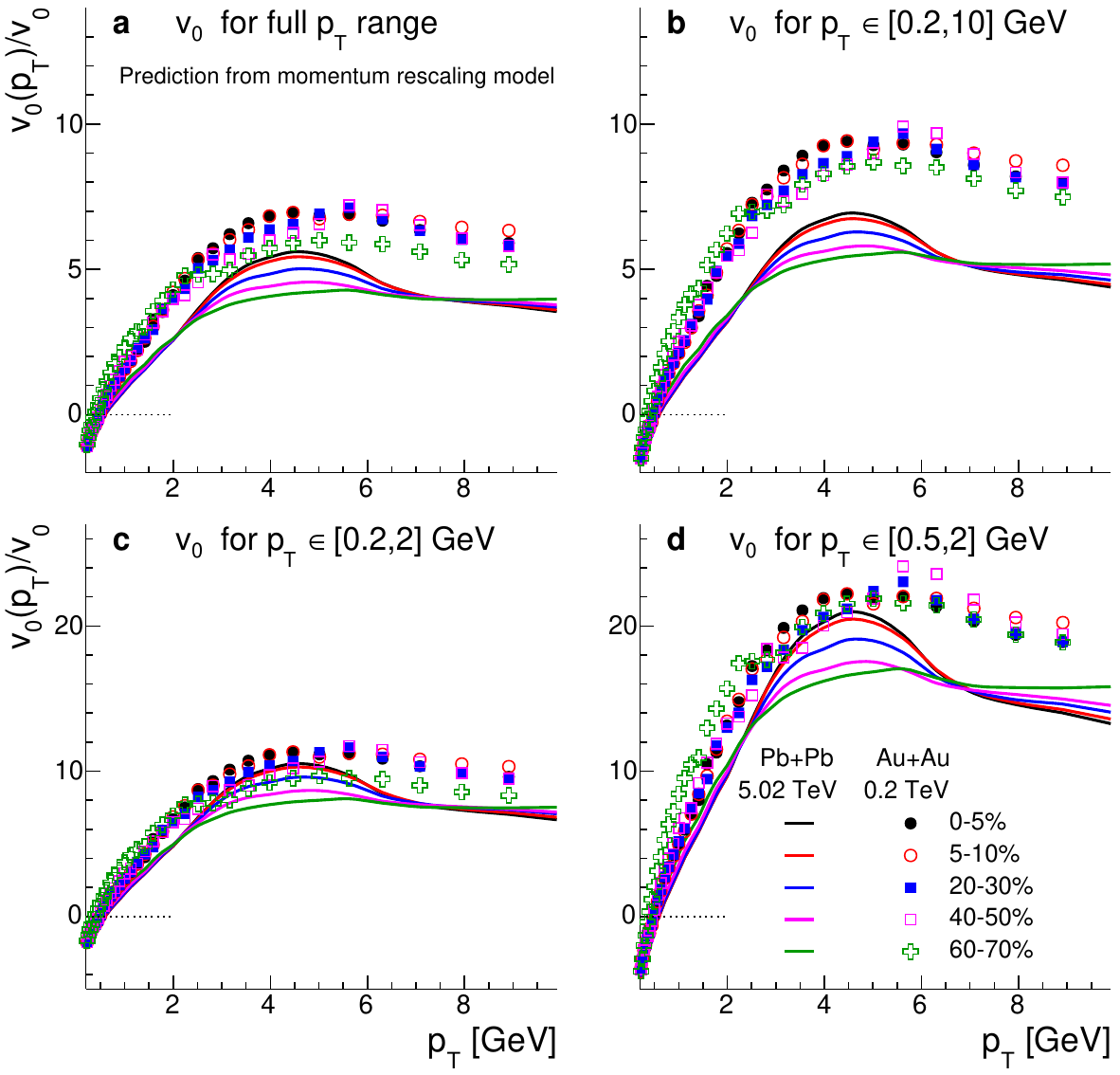}
\vspace*{-0.2cm}
    \caption{\textbf{Predicted $v_0(\pT)/v_0$ at RHIC and LHC energies based on momentum rescaling model.}
The $v_0(\pT)/v_0$ calculated from charged hadron spectra via Eq.~\eqref{eq:5} for Pb+Pb collisions at $\snn = 5.02$ TeV (solid lines, LHC) and Au+Au collisions at $\snn = 0.2$ TeV (markers, RHIC). Panels show results normalized to $v_0$ calculated in different reference $\pT$ ranges: (\textbf{a}) full $\pT$ range, (\textbf{b}) $0.2 < \pT < 10$ GeV range, (\textbf{c}) $0.2 < \pT < 2$ GeV range, and (\textbf{d}) $0.5 < \pT < 2$ GeV range.}
\label{fig:4}
\vspace*{-0.2cm}
\end{figure}

{\bf Prediction at RHIC energy.}
Having demonstrated the framework's success in describing LHC data, we extend our predictions to RHIC energies using published spectra from $pp$ and Au+Au collisions at $\snn=0.2$ TeV~\cite{STAR:2003fka,PHENIX:2003djd}. Figure~\ref{fig:4} presents these predictions alongside LHC results, with each panel showing a different reference momentum range $\pTa$ for normalization.

The predicted $v_0(\pT)/v_0$ patterns at RHIC exhibit several systematic differences from LHC. Most notably, the RHIC predictions are systematically higher across all $\pT$ regions. At low $\pT$ ($<2$ GeV), this enhancement arises from the smaller average momentum $\lr{[\pT]}$ at RHIC, which according to Eq.~\eqref{eq:5a} directly increases $v_0(\pT)/v_0$. This trend was also observed in hydrodynamic calculations~\cite{Jahan:2025cbp,Du:2025dpu}. At high $\pT$ ($>5$ GeV), the enhancement reflects the steeper power-law tail at RHIC, characterized by a larger spectral index $m$ (yielding $m-1=7$ versus 5 at LHC according to Eq.~\eqref{eq:5b}). This is very different from jet quenching, which suppresses the spectra at both energies in the 5--10 GeV range by a similar amount, yet our predictions show enhanced $v_0(\pT)/v_0$ at RHIC.

The apparently weaker centrality dependence at RHIC has a simple origin: the smaller difference between the hydrodynamic baseline at around $\pT\sim 2$--3 GeV (Eq.~\eqref{eq:5a}) and the power-law limit (Eq.~\eqref{eq:5b}) at RHIC compresses the range of variation. The choice of reference range significantly influences the observed energy dependence. The range $\pTa\in[0.2,10]$ GeV maximizes the RHIC-LHC difference, while $\pTa\in[0.2,2]$ GeV minimizes it. This sensitivity arises because radial flow fluctuations are distributed differently across $\pT$ at the two energies. As demonstrated in Appendix~\ref{app:1} (Fig.~\ref{fig:5}), RHIC concentrates a larger fraction of total momentum variance at low $\pT$ due to its steeper spectra, affecting the normalization factor $v_0^A$ for any given reference range.

These predictions establish an important baseline: substantial variations in $v_0(\pT)/v_0$ between RHIC and LHC can arise purely from spectral shape differences, independent of any dynamical physics. This has important implications for interpreting future measurements at low $\snn$.

{\bf Discussion.} The momentum rescaling framework developed here directly addresses the interpretational challenges raised in the introduction and provides new insights for future studies.

First, our analysis reveals that the characteristic rise-and-fall pattern of $v_0(\pT)$ emerges primarily from the $\pT$-dependence of the spectral index. At low $\pT$, the steep exponential spectrum produces a linear rise of $v_0(\pT)$. As $\pT$ increases, the transition from exponential to power-law behavior creates the peak around 3--4 GeV. At high $\pT$, the power-law asymptote leads to the gradual decline. Critically, the substantial energy dependence predicted between RHIC and LHC differs fundamentally from the jet quenching scenario, which predicts similar suppression patterns at both energies.

Second, the sensitivity of $v_0(\pT)$ to spectral shape imposes stringent requirements on modeling. Any hydrodynamic model attempting to extract bulk viscosity or other medium properties from $v_0(\pT)$ must first accurately reproduce the particle spectra. Otherwise, variations in $v_0(\pT)$ when varying viscosity could reflect changes in the predicted spectra rather than genuine viscosity effects. The non-equilibrium corrections during Cooper-Frye hadronization (so-called $\delta f$ corrections)~\cite{Ryu:2017qzn,Du:2025dpu} are also important, as they directly modify spectral shape and thus influence $v_0(\pT)$ predictions.

Third, the significant energy dependence of the kinematic baseline shown in Fig.~\ref{fig:4} cautions against naive comparisons of $v_0(\pT)/v_0$ between RHIC and LHC. Observed differences could simply reflect spectral shape variations rather than fundamental changes in QGP properties. Instead, the dynamical component $g(\pT)$ provides a more robust basis for comparison. Since the QGP properties at RHIC and LHC are expected to be qualitatively similar, similar $g(\pT)$ patterns at both energies would support the universality of the underlying dynamics, while confirming that spectral shape variations dominate the apparent energy dependence of $v_0(\pT)/v_0$.

{\bf Future extensions.} The momentum rescaling framework provides a foundation for investigating radial flow fluctuations beyond $v_0(\pT)$. Two extensions to other experimental observables are discussed here.

{\it Two-particle correlations.} The framework extends naturally to two-particle correlations~\cite{Schenke:2020uqq}, where multiplicity fluctuations in different momentum ranges $\pa$ and $\pb$ are correlated:
\begin{align}\label{eq:13}
V_0(\pa,\pb)=\frac{\lr{\delta n(\pa)\delta n(\pb) }}{\lr{n(\pa)}\lr{n(\pb)}}\;.
\end{align}
Collective radial flow implies factorization: $V_0(\pa,\pb)= v_0(\pa)v_0(\pb)$. Within our model, this yields a testable prediction:
\small{\begin{align}\label{eq:14}
V_0(\pa,\pb) \approx \left(\frac{d\ln \lr{n}}{d\ln\pT} (\pa)+1\right)\left(\frac{d\ln \lr{n}}{d\ln\pT} (\pb)+1\right)\;.
\end{align}}\normalsize
Deviations from this spectral-shape prediction would reveal momentum-dependent decorrelation arising from subleading flow modes~\cite{Mazeliauskas:2015efa}, providing a clean test of the single-mode rescaling assumption.

{\it Higher-order fluctuations.} The approach generalizes to higher-order spectral cumulants~\cite{Bhatta:2025oyp}:
\begin{align}\nonumber
&v_{0}\{3\}(\pT)^3 = \lr{(\delta n(\pT))^3}/\lr{n(\pT)}^3 \;,\\\nonumber
&v_{0}\{4\}(\pT)^4 = \frac{\lr{(\delta n(\pT))^4}-3\lr{(\delta n(\pT))^2}^2}{\lr{n(\pT)}^4}\;,\\\label{eq:15}
&...
\end{align}
Under global rescaling, all cumulants exhibit identical $\pT$ dependence (Eq.~\eqref{eq:4}):
\begin{align}\label{eq:16}
\frac{v_0\{k\}(\pT)}{v_0\{k\}} \approx -\frac{d\ln \lr{n(\pT)}}{d\ln\pT}-1\;.
\end{align}
This prediction extends to multi-particle correlations with each particle from a different $\pT$ range:
\begin{align}
V_0\{k\}\left(p_{\mathrm{T} 1}, p_{\mathrm{T} 2,...}\right)\approx \overset{k}{\underset{i=1}{\prod}} \left(-\frac{d \ln \lr{n}}{d\ln\pT}(p_{\mathrm{T} i})-1\right)\;.
\end{align}
Deviations from these baselines would signal non-Gaussian dynamics or $\pT$-dependent correlations beyond simple $\pT$ rescaling. Such measurements could probe the pdf of radial flow fluctuations and test the Gaussian approximation underlying hydrodynamic models.

{\bf Summary and outlook.}
We introduced a momentum rescaling framework that factorizes $v_0(\pT)$ into kinematic and dynamical components: $v_0(\pT)/v_0 = -[d\ln\langle n(\pT)\rangle/d\ln\pT + 1] \times g(\pT)$. The first factor, determined purely by spectral shape, already captures the characteristic rise-and-fall pattern. The second factor, $g(\pT)$, isolates genuine $\pT$-dependent dynamics beyond global momentum rescaling.

Analysis of ATLAS data reveals that $g(\pT)$ deviates from unity by 20--40\%. These deviations are strongest in central collisions and systematically weaken toward peripheral collisions, indicating their connection to medium effects. Predictions for RHIC energies demonstrate that spectral shape differences alone generate substantial energy-dependent variations in $v_0(\pT)/v_0$, establishing a clear baseline for identifying genuine dynamical effects in future beam-energy-scan measurements.

This framework provides a natural link between experimental observables and theoretical predictions. Dynamical models, such as blast-wave parameterizations, viscous hydrodynamics, or models including jet quenching, can make specific predictions for $g(\pT)$ that directly encode its physics content. Future measurements of $g(\pT)$ for identified particles could test the universality of radial flow. This framework also extends naturally to higher-order fluctuations (Eq.~\eqref{eq:16}), where deviations from factorization signal $\pT$-dependent non-Gaussian dynamics. 

The momentum rescaling principle may extend beyond radial flow. Anisotropic flow $v_n(\pT)$ could potentially be understood through azimuthally-dependent momentum rescaling, enabling investigation of correlations between isotropic and anisotropic collective phenomena such as $\langle v_n(\pT)^2 v_0(\pT)\rangle$~\cite{Parida:2025eqv}. This connection warrants future exploration. More broadly, this work demonstrates that disentangling kinematic from dynamical effects is essential for extracting robust insights about QGP properties from fluctuation observables.

This work is supported by DOE Research Grant Number DE-SC0024602. The author acknowledges valuable discussions with Derek Teaney, Somadutta Bhatta, Zhengxi Yan, and Giuliano Giacalone. The author acknowledges the use of AI-based Glaude, for grammar and textural improvements.

\section*{End Matter}
\subsection{Relating the integral $v_0$ in different $\pT$ ranges}\label{app:1}
Different experiments measure $v_0=\sqrt{\lr{(\delta [\pT])^2}}/\lr{[\pT]}$ over different $\pT$ ranges due to varying detector acceptances and analysis strategies. STAR and ALICE typically use $\pT>0.2$ GeV, CMS employs $\pT>0.3$ GeV, while ATLAS uses $\pT>0.5$ GeV. Though these measurements probe the same underlying radial flow fluctuations, their numerical values differ substantially due to $\pT$-range dependence. Understanding this dependence is crucial for comparing results across experiments and collision energies.

We quantify this effect using the acceptance factor $C_A$ introduced in Ref.~\cite{Parida:2024ckk}, which relates measurements within a specific range $\pT\in A=[p_{\mathrm{T}}^{\mathrm{min}},p_{\mathrm{T}}^{\mathrm{max}}]$ to the full-acceptance value via Eq.~\eqref{eq:8}. This factor encodes how spectral shape and $v_0(\pT)$ weight different momentum regions.

\begin{figure}[h!]
    \centering
\vspace*{-0.4cm}
 \includegraphics[width=1\linewidth]{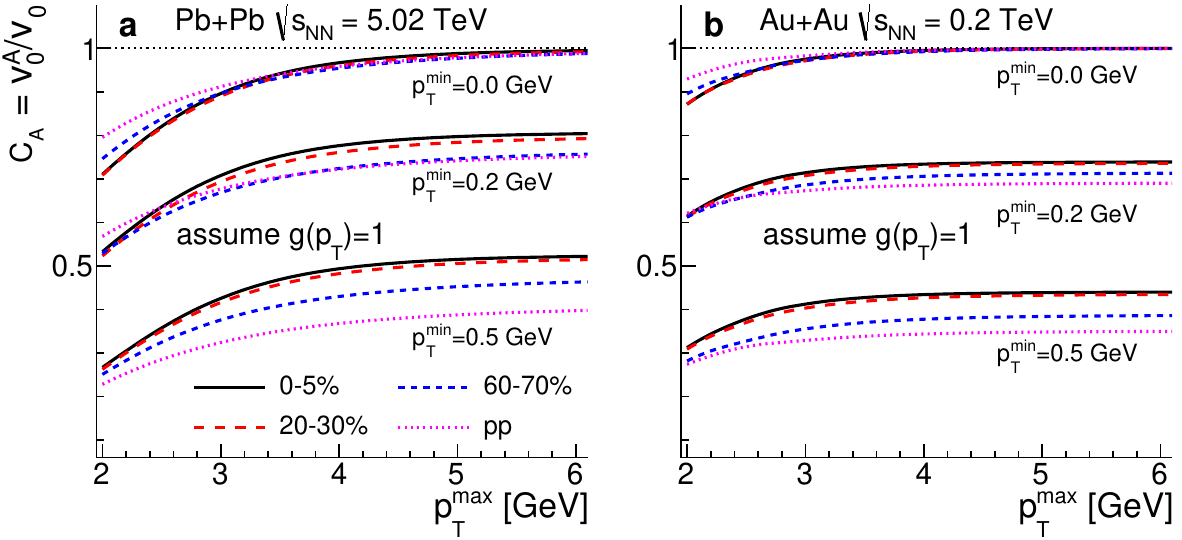}
    \caption{\textbf{Acceptance factor versus momentum range boundaries.} The acceptance factor $C_A = v_0^A/v_0$ (Eq.~\eqref{eq:8}) versus $p_{\mathrm{T}}^{\mathrm{max}}$ for different $p_{\mathrm{T}}^{\mathrm{min}}$ values (labeled). Results are shown for Pb+Pb at $\snn = 5.02$ TeV (\textbf{a}) and Au+Au at $\snn = 0.2$ TeV (\textbf{b}), with four centrality classes per group. The factor increases with $p_{\mathrm{T}}^{\mathrm{max}}$ and saturates differently at RHIC versus LHC due to spectral shape differences.}
    \label{fig:5}
\end{figure}

Figure~\ref{fig:5} presents predictions for $C_A$ as a function of acceptance boundaries, revealing three key features: (1) $C_A$ decreases in peripheral collisions for larger $p_{\mathrm{T}}^{\mathrm{max}}$, reflecting their flatter spectra; (2) RHIC shows systematically larger $C_A$ than LHC due to steeper spectra at lower energies; (3) The acceptance factor saturates at $p_{\mathrm{T}}^{\mathrm{max}} > 5$--6 GeV for LHC but earlier ($\sim$3--4 GeV) for RHIC.

These trends arise because $C_A$ weights contributions from both the local yield $\lr{n(\pT)}$ and the local $v_0(\pT)$ value (Eq.~\eqref{eq:8}). In peripheral collisions, the flatter spectrum reduces the relative contribution from low-$\pT$ particles where $v_0(\pT)$ is negative, thereby decreasing $C_A$.

\begin{figure}[h!]
    \centering
 \includegraphics[width=1\linewidth]{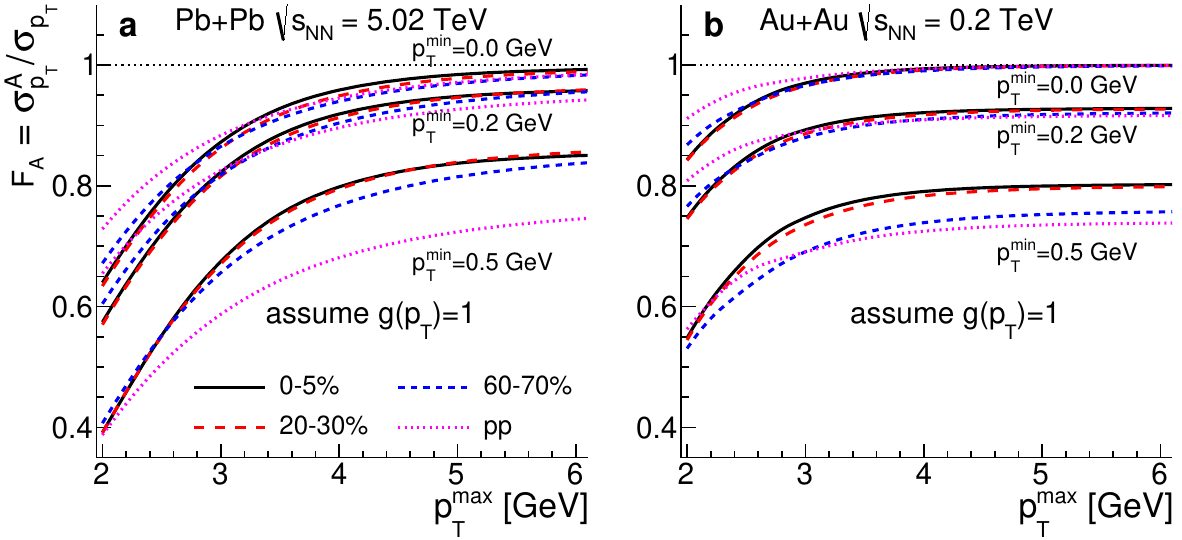}
\vspace*{-0.4cm}
    \caption{\textbf{Momentum variance fraction versus momentum range boundaries.} The variance fraction $F_A$ (Eq.~\eqref{eq:20}) versus $p_{\mathrm{T}}^{\mathrm{max}}$ for different $p_{\mathrm{T}}^{\mathrm{min}}$ values (labeled), showing how total momentum fluctuations are distributed across $\pT$. Results are shown for Pb+Pb at $\snn = 5.02$ TeV (\textbf{a}) and Au+Au at $\snn = 0.2$ TeV (\textbf{b}), with four centrality classes per group. RHIC shows higher fractions at low $\pT$ due to steeper spectra, with saturation reflecting the momentum reach of collective flow.}
    \label{fig:6}
\end{figure}

Since $v_0$ depends strongly on $\lr{[\pT]}$, which varies considerably with $\pT$-range, we introduce a complementary observable removing this dependence:
\begin{align}\label{eq:20}
F_A = C_A \frac{\lr{[\pT]}}{\lr{[\pT]}_A} = \frac{\sqrt{\lr{(\delta [\pT])^2}_A}}{\sqrt{\lr{(\delta [\pT])^2}}}\;.
\end{align}
This quantity directly measures the fraction of total momentum variance contained within range $A$.

Figure~\ref{fig:6} shows that $F_A$ is substantially larger than $C_A$ and increases at lower $\snn$. The reference range $0.5<\pT<2$ GeV captures approximately 40\% of total fluctuations at LHC but 55\% at RHIC, reflecting the concentration of particle production at lower momenta for smaller $\snn$. The saturation of $F_A$ with $p_{\mathrm{T}}^{\mathrm{max}}$ provides a direct measure of the momentum reach of collective radial flow: this influence extends to 5--6 GeV at LHC but only 3--4 GeV at RHIC.

\subsection{Validity of linear approximation for $v_0(\pT)/v_0$}
The formula for $v_0(\pT)$ in Eq.~\eqref{eq:5} relies on small momentum fluctuations (or $v_0\ll1$) and first-order Taylor expansion. This linear approximation breaks down for large $v_0$ values relevant to peripheral collisions and small systems. To establish the validity range and illustrate this limitation, we examine an exponential spectrum:
\begin{align}\label{eq:21}
\lr{n(\pT)} \propto \frac{\pT}{[\pT]^2} \exp\left(-\frac{2\pT}{[\pT]}\right)\;.
\end{align} 
For this spectrum, the exact result in the small-$v_0$ limit is~\cite{Schenke:2020uqq}:
\begin{align}\label{eq:22}
v_0(\pT) = 2 v_0 \left(\frac{\pT}{\lr{[\pT]}}-1\right).
\end{align}

\begin{figure}[!htbp]
\centering
\includegraphics[width=0.8\linewidth]{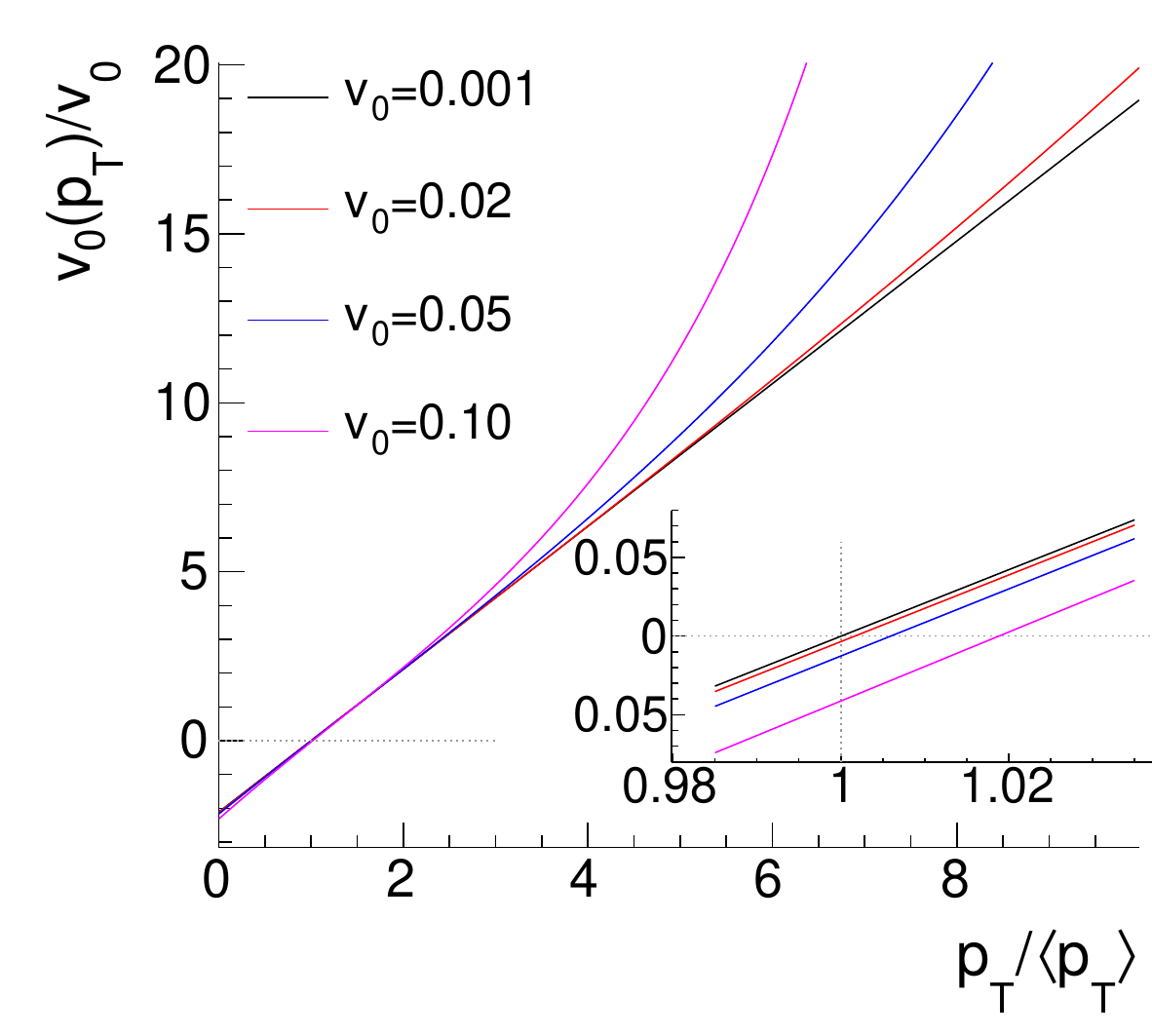}
 \caption{\textbf{Breakdown of linear approximation for large fluctuations.} Exact $v_0(\pT)/v_0$ for exponential spectra (Eq.~\eqref{eq:21}) at different $v_0$ magnitudes, showing systematic deviations from the linear prediction (Eq.~\eqref{eq:22}) as $v_0$ increases. The zero-crossing point shifts to higher $\pT$ (inset), and shape distortions become significant for $v_0>0.05$, indicating breakdown of Eq.~\eqref{eq:5} requiring nonlinear corrections in peripheral collisions and small systems.}
\label{fig:7}
\end{figure}

Experimental measurements span a wide range of $v_0$ values. In Au+Au (RHIC)~\cite{STAR:2005vxr,Adam:2019rsf,STAR:2024wgy,STAR:2025elk} and Pb+Pb (LHC)~\cite{ALICE:2014gvd,ALICE:2024apz,ATLAS:2024jvf}, $v_0$ ranges from $\sim$0.01 (central) to $>$0.05 (peripheral). Small systems show even larger values: 0.05--0.14 in $pp$ collisions~\cite{ALICE:2014gvd}, with generally higher values at RHIC due to smaller $\lr{[\pT]}$~\cite{Adam:2019rsf}. Crucially, most measurements use restricted $\pT$ ranges (typically 0.2--2 GeV), while full-acceptance $v_0$ values could be $\sim$2 times larger (Fig.~\ref{fig:5}). This implies total $v_0$ values could reach 0.02 in central, 0.05 in mid-central, and $>$0.1 in peripheral collisions. 

Figure~\ref{fig:7} quantifies deviations from the linear approximation (Eq.~\eqref{eq:22}) for different $v_0$ magnitudes, revealing:
\begin{itemize}
\item For $v_0\lesssim0.02$ (central collisions): deviations remain below 5\%.
\item For $v_0=0.05$ (mid-peripheral): 10\% deviations appear at $\pT\sim 4$ GeV.
\item For $v_0=0.1$ (peripheral/small systems): deviations reach 40\% at high $\pT$.
\item The zero-crossing point shifts systematically to larger $\pT$ with increasing $v_0$.
\end{itemize}

These results have important implications. The linear framework remains accurate for central and mid-central collisions ($v_0 \lesssim 0.05$), while peripheral collisions require nonlinear corrections, particularly at high $\pT$. In extreme cases ($v_0>0.1$), the same spectrum can produce qualitatively different $v_0(\pT)/v_0$ shapes purely from fluctuation magnitude. Future work should develop systematic nonlinear expansions or numerical approaches for proper interpretation of $v_0(\pT)$ in these regimes, particularly important for small systems where collective behavior remains debated.

\bibliography{../../v0pt}{}
\bibliographystyle{apsrev4-1}
\end{document}